\documentclass[amssymb,prb,twocolumn,superscriptaddress,floats,showkeys,amsmath]{revtex4-1}

\usepackage{graphicx}
\usepackage{epstopdf}
\usepackage{hyperref} 
\hypersetup{colorlinks,citecolor=blue, filecolor=blue ,linkcolor=blue , urlcolor=blue, pdftex}
\usepackage[euler]{textgreek}

\usepackage[labeled,resetlabels]{multibib}
\newcites{SI}{SupI}

\def \FUW{Institute of Experimental Physics, Faculty of Physics, University of Warsaw, ul. Pasteura 5, 02-093 Warsaw, Poland}
\def \LNCMI{Laboratoire National des Champs Magn\'etiques Intenses, CNRS-UGA-UPS-INSA-EMFL, 25, avenue des Martyrs, 38042 Grenoble, France} 
\def \Prague{Department of Condensed Matter Physics, Faculty of Mathematics and Physics, Charles University in Prague, Ke Karlovu 5, Praha 2 CZ-121 16, Czech Republic}
\def \Brno{Central European Institute of Technology, Brno University of Technology,  Purky\v{n}ova 656/123, 612 00 Brno, Czech Republic}
\def \Wroclaw{Department of Experimental Physics, Faculty of Fundamental Problems of Technology, Wroc\l{}aw University of Science and Technology, ul. Wybrze\.ze Wyspia\'nskiego 27, 50-370 Wroc\l{}aw, Poland}
\def \Kenji{Research Center for Functional Materials, National Institute for Materials Science, 1-1 Namiki, Tsukuba 305-0044, Japan}
\def \Takashi{International Center for Materials Nanoarchitectonics, National Institute for Materials Science, 1-1 Namiki, Tsukuba 305-0044, Japan}

\begin{document}

\title{Neutral and charged dark excitons in monolayer WS$_2$}

\author{M. Zinkiewicz}\email{malgorzata.zinkiewicz@fuw.edu.pl}\affiliation{\FUW}
\author{A. O. Slobodeniuk}\affiliation{\Prague}
\author{T. Kazimierczuk}\affiliation{\FUW}
\author{P. Kapu\'sci\'nski}\affiliation{\LNCMI}\affiliation{\Wroclaw{}}
\author{K.~Oreszczuk}\affiliation{\FUW}
\author{M.~Grzeszczyk}\affiliation{\FUW}
\author{M. Bartos}\affiliation{\LNCMI}\affiliation{\Brno}
\author{K. Nogajewski}\affiliation{\FUW}
\author{K.~Watanabe}\affiliation{\Kenji}
\author{T.~Taniguchi}\affiliation{\Takashi}
\author{C. Faugeras}\affiliation{\LNCMI}
\author{P. Kossacki}\affiliation{\FUW}
\author{M. Potemski}\affiliation{\FUW}\affiliation{\LNCMI}
\author{A. Babi\'nski}\affiliation{\FUW}
\author{M. R. Molas}\email{maciej.molas@fuw.edu.pl}\affiliation{\FUW}

\begin{abstract}

Low temperature and polarization resolved magneto-photoluminescence experiments are used to investigate the properties of dark excitons and dark trions in a monolayer of WS$_2$ encapsulated in  hexagonal BN (hBN). We find that this system is an $n$-type doped semiconductor and that dark trions dominate the emission spectrum. In line with previous studies on WSe$_2$, we identify the Coulomb exchange interaction coupled neutral dark and grey excitons through their polarization properties, while an analogous effect is not observed for dark trions. Applying the magnetic field in both perpendicular and parallel configurations with respect to the monolayer plane, we determine the g-factor of dark trions to be $g\sim$-8.6. Their decay rate is close to 0.5 ns, more than 2 orders of magnitude longer than that of bright excitons.


\end{abstract}

\maketitle

\section{Introduction \label{sec:Intro}}

Monolayers (MLs) of semiconducting transition metal dichalcogenides (S-TMDs) MX$_2$ where M=Mo or W and X=S, Se or Te, are direct band gap semiconductors with the minima (maxima) of conduction (valence) band located at the inequivalent K$^+$ and K$^-$ points of their hexagonal Brillouin zone (BZ) \cite{Koperski2017, Wang2018}. The strong spin-orbit interaction in the crystal lifts the spin-degeneracy of the bands, which leads, in particular, to the splitting of the valence ($\Delta_v$) and the conduction ($\Delta_c$) bands. While the former splitting is of the order of few hundreds of meV, the latter equals few tens of meV only. Moreover, the $\Delta_c$ splitting can be positive or negative. As a result two subgroups of MLs can be distinguished: $bright$ (the excitonic ground state is optically active or bright) and $darkish$ (the excitonic ground state is optically inactive or dark). It is well established that WS$_2$ and WSe$_2$ MLs belong to the group of darkish materials \cite{Molas2017,Zhang2017}, while MLs of MoSe$_2$ and MoTe$_2$ belong to the family of bright materials \cite{Molas2017, Koperski2017, Robert2020}. The assignment of the MoS$_2$ ML is still under debate as theoretical predictions and experimental results are contradictory \cite{Kormanyos2015, Molas2017, Robert2020}. 

\begin{figure}[b]
	\centering
	\includegraphics[width=0.5\linewidth]{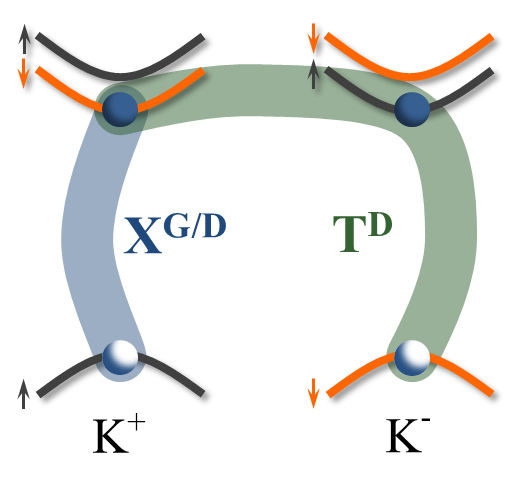}%
	\caption{Schematic illustration of possible configurations for the neutral (blue shade) and the negatively charged dark (green shade) excitons located at the K$^+$ and K$^-$ valleys, respectively. The grey (orange) curves indicate the spin-up (spin-down) subbands. The electrons (holes) in the conduction (valence) band are represented by blue (white) circles.}
	\label{fig:XDTD}
\end{figure}

The complex electronic structure of S-TMD MLs results in a wide variety of excitonic complexes, which can be formed from carriers at the vicinity of the conduction (CB) and valence band (VB) extrema. Their studies have been largely facilitated by the efforts to improve their optical quality, in particular by encapsultaing MLs of S-TMDs in thin layers of hexagonal BN (hBN). In particular, this improvement allowed to investigate several dark excitonic complexes \cite{Robert2017, Barbone2018, Chen2018, Li2018, Paur2019, Molas2019Dark,Li2019Trion, Liu2019Replica, Li2019Replica, Paur2019, Li2019Momentum, arora2019dark, he2020valley, Robert2020}. Among them there are spin- or momentum-forbidden dark excitons, which can not recombine optically due to spin or momentum conservation laws. Moreover, these complexes can be neutral and charged. Fig.~\ref{fig:XDTD} illustrates schematically the neutral and negatively charged spin-forbidden dark excitons in darkish monolayer ($i.e.$ WS$_2$ or WSe$_2$). The neutral dark exciton at K$^+$ valley is composed of an electron from the lowest-lying level of CB and a hole from the highest-lying level of VB. The negative trion at the K$^-$ point is formed similarly by the energetically lowest electron-hole ($e−h$) pair and an extra electron located in the K$^+$ valley. It was proposed theoretically and demonstrated experimentally that the spin-forbidden neutral dark excitons exhibit a double (fine) structure comprising so-called grey (X$^\textrm{G}$) and dark (X$^\textrm{D}$) complexes. These complexes are characterized by the out-of-plane and zero excitonic dipole momenta \cite{Slobodeniuk2016, Robert2017, Molas2019Dark}, respectively. Although substantial efforts have been made to study dark excitons in WSe$_2$ MLs \cite{WangMarie2017, Robert2017, Liu2019, Molas2019Dark, Li2019Trion, Liu2019Replica, Li2019Replica, Paur2019, Li2019Momentum, arora2019dark, he2020valley}, their properties in WS$_2$ are still far from complete understanding \cite{WangMarie2017, Paur2019, arora2019dark}. For example, the reported energy difference between the bright and dark excitonic emission is not well determined as it varies significantly (55~meV \cite{WangMarie2017} vs. 46~meV \cite{Paur2019}). This motivates our addressing properties of dark states in WS$_2$ ML.

In this work, we use polarization resolved photoluminescence spectroscopy with an applied magnetic field to investigate dark excitonic complexes through the magnetic brightening effect in a high quality and naturally $n$-doped WS$_2$ ML encapsulated in thin hBN layers. The magnetic field has been applied in different geometries: perpendicular, parallel or at 45$^\circ$ with respect to the monolayer plane. The emissions from both the neutral and the charged exciton complexes are activated by the in-plane magnetic field component. The double structure of the neutral dark exciton (grey and dark excitons) is observed while the dark trions do not present any fine structure. Our study also shows that the magnetic brightening of the dark excitons and dark trions depends on the carrier concentration.

\section{Results \label{results}}

\begin{figure}[!t]
	\centering
	\includegraphics[width=1\linewidth]{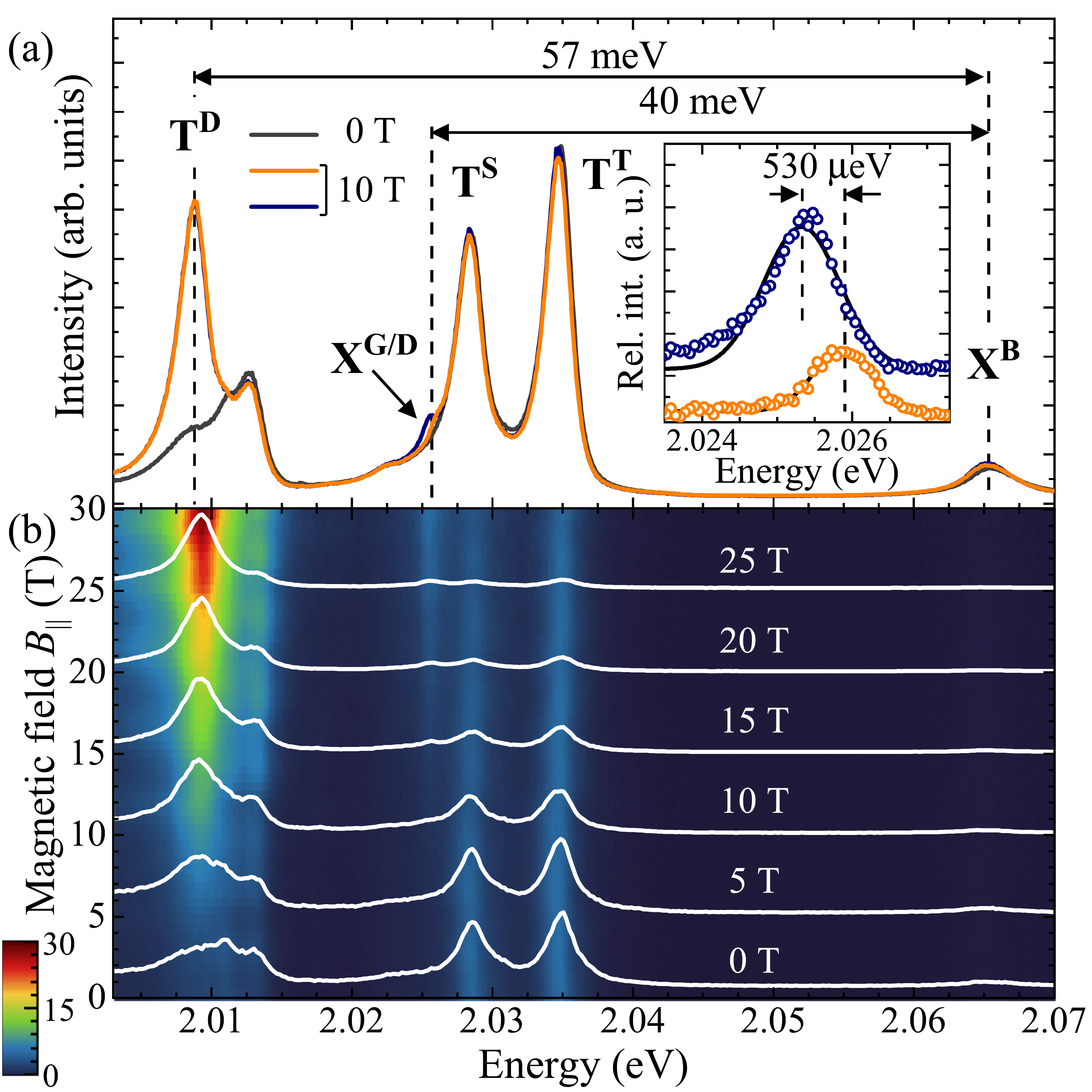}%
	\caption{(a) Low-temperature PL spectra measured on a WS$_2$ ML encapsulated in hBN flakes at zero magnetic field and at $B_{||}$=10 T. Grey curve represents unpolarized detection, while the blue and orange curves correspond to the detection of two linear polarizations aligned parallel and perpendicular to the direction of magnetic field, respectively. The inset shows the relative intensities of the grey and dark exciton emissions defined as (PL$_{B\textrm{=10 T}}$ - PL$_{B\textrm{=0 T}}$)/PL$_{B\textrm{=0 T}}$. (b)~False-color map of the PL response  as a function of $B_{||}$ from zero field to 30~T. The intensity scale is normalized to the T$^\textrm{T}$ intensity to remove the  Faraday effect affecting the experimental data. White curves superimposed on the map represent the PL spectra normalized to the most intense peaks recorded at selected values of $B_{||}$.}
	\label{fig:mapa}
\end{figure}

Fig.~\ref{fig:mapa} illustrates the brightening of neutral and charged dark excitons in a monolayer of WS$_2$ encapsulated in hBN by an in-plane magnetic field. The zero-field PL spectrum is composed of several emission lines. Based on the previous reports, \cite{Vaclavkova2018,nagler2018,Jadczak2019,Paur2019,Molas2019Spectrum} three peaks can be ascribed unquestionably to a bright exciton (X$^\textrm{B}$) and singlet (T$^\textrm{S}$) and triplet (T$^\textrm{T}$) states of negative trions.  The application of an in-plane magnetic field $B_{||}$ results in the appearance of three additional lines, labelled X$^\textrm{G}$, X$^\textrm{D}$ and T$^\textrm{D}$. This can be appreciated in Fig.~\ref{fig:mapa}, which displays the spectrum measured at $B_{||}$=10~T. The X$^\textrm{G}$ and X$^\textrm{D}$ peaks correspond to the dark and grey states of the neutral exciton, while the T$^\textrm{D}$ peak is related to a dark state of the negative trion. Moreover, the detailed analysis of the X$^\textrm{D/G}$ line, shown in the inset to Fig.~\ref{fig:mapa}(a), exhibits its fine structure, in line with previous studies on WSe$_2$ MLs \cite{Courtade2017, Molas2019Dark}. These results can be summarized as follows: (i) the X$^\textrm{D/G}$ line is red-shifted by 40 meV from the X$^\textrm{B}$ peak; (ii) the energy separation between the X$^\textrm{G}$ and X$^\textrm{D}$ emissions $\delta$=530~$\mu$eV; (iii) the T$^\textrm{D}$ resonance is red-shifted by 57 meV from the X$^\textrm{B}$ one. Surprisingly, the energy separation between the negatively charged/neutral dark and bright neutral exciton in the studied WS$_2$ ML, which amounts to 57~meV/40~meV, are very similar to the corresponding values observed in a WSe$_2$ ML encapsulated in the same dielectric environment ($\sim$57~meV/$\sim$40~meV) \cite{WangMarie2017, Robert2017, Liu2019, Molas2019Dark, Li2019Trion, Liu2019Replica, Li2019Replica, Paur2019, Li2019Momentum,he2020valley}. The obtained value for the fine structure splitting of the neutral dark-grey exciton of about 530~$\mu$eV in the studied WS$_2$ ML is also very similar to that observed in WSe$_2$ (660~$\mu$eV) \cite{Robert2017,Molas2019Dark}.

\begin{figure}[t]
	\centering
	\includegraphics[width=1\linewidth]{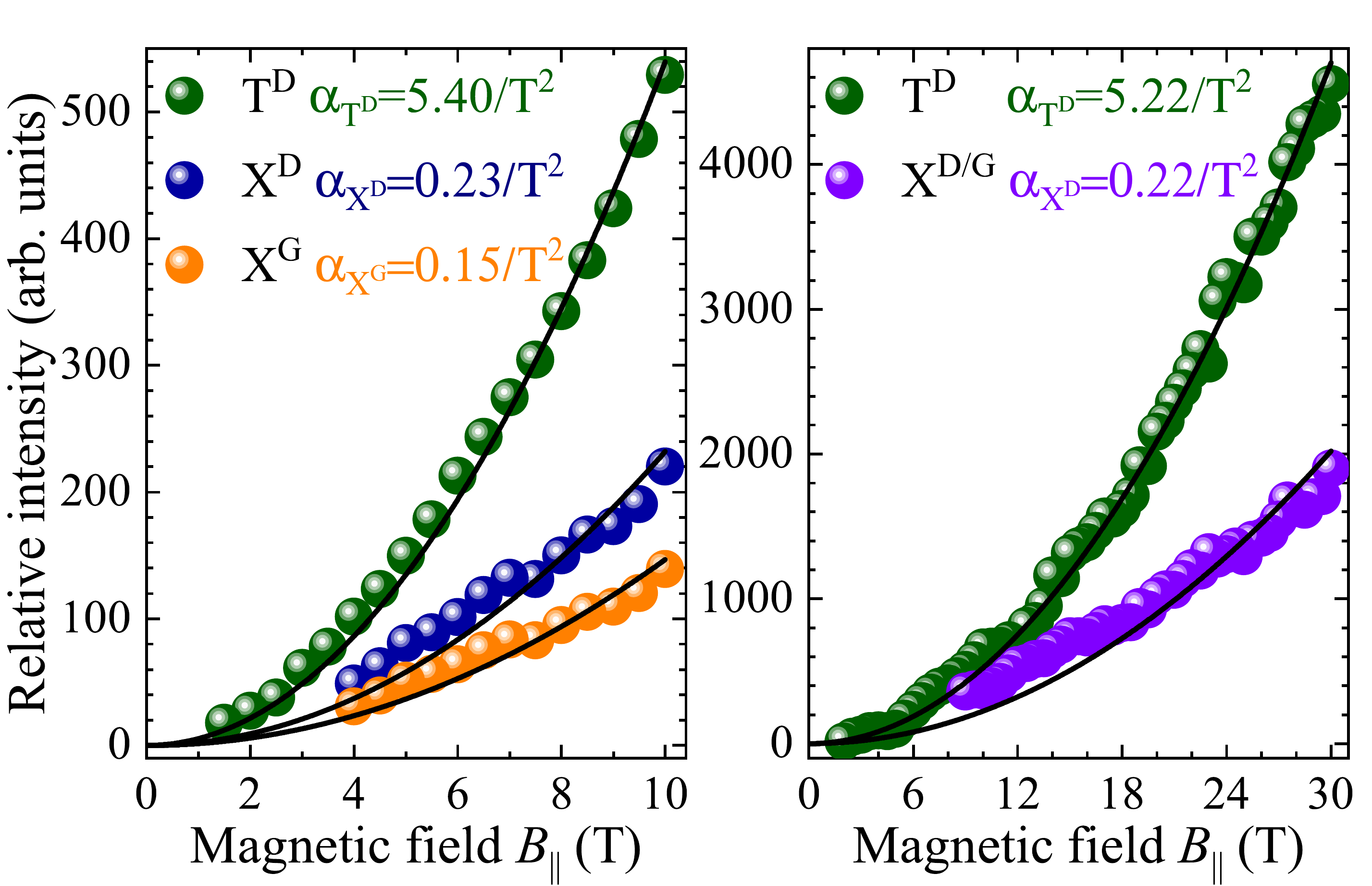}%
	\caption{The effect of the in-plane magnetic field $B_{||}$ on the integrated relative intensities of the charged and neutral dark excitons measured in magnetic fields up to (a) 10 T and (b) 30 T. The spectra were detected in (a) two linear polarizations for X$^\textrm{G}$ and X$^\textrm{D}$ emissions (see Fig.~\ref{fig:mapa}(a) for details), and otherwise detection was unpolarized. Note that the X$^\textrm{D/G}$ intensities in both panels were multiplied by factor 5 for clarity. The solid black curves represent quadratic fits.}
	\label{fig:Fig2}
\end{figure}

Having identified the dark excitons, we now investigate these complexes at higher magnetic fields (up to 30 T) but with unpolarized optical detection. In fact, as can be seen in Fig.~\ref{fig:mapa}(b), the application of the in-plane magnetic field $B_{||}$ up to 30~T leads to a strong brightening effect, which is significantly different for the neutral and charged dark excitons. With increasing $B_{||}$, the intensity of the dark trion increases substantially, while the intensity of the bright exciton and trions stays practically unchanged. As can be seen in Fig.~\ref{fig:mapa}(b), the emission of the neutral grey-dark exciton doublet is much smaller as compared to dark trion even at highest magnetic fields ($B$=30~T). The $B_{||}$ evolution of the intensity of dark excitons is expected to be quadratic $I=\alpha B_{||}^2$~~\cite{Molas2017, Zhang2017, Molas2019Dark}. Note that we neglect the zero-field intensity of the grey components of the neutral and charged complexes, as we were not able to resolve it at zero magnetic field. Fig.~\ref{fig:Fig2} demonstrates the $B_{||}^2$ dependence of the relative intensities of the charged and neutral dark excitons in magnetic fields up to 10~T and 30~T, which are accompanied with the quadratic fits. The integrated relative intensities of the T$^\textrm{D}$, X$^\textrm{G}$ and X$^\textrm{D}$ lines were obtained by fitting the relative spectra defined as (PL$_{B\neq \textrm{0 T}}$ - PL$_{B\textrm{=0 T}}$)/$I_{\textrm{X}^\textrm{B}}(B)$ using Lorentzian functions. The PL$_{B\neq \textrm{0 T}}$ and PL$_{B\textrm{=0 T}}$ are correspondingly photoluminescence spectra measured at non-zero and zero magnetic fields, while $I_{\textrm{X}^\textrm{B}}(B)$ represents the integrated intensity of the bright neutral exciton as a function of magnetic fields. Note that the division by the $I_{\textrm{X}^\textrm{B}}(B)$ parameter allows us to eliminate variation of the signal intensity during measurements, $e.g.$  the measured signal in 30~T setup is affected by the Faraday effect. The fitted $\alpha$ parameters for three analysed excitons observed in both experimental setups are coherent within experimental error. However, as it is shown in Figs~\ref{fig:Fig2}(a) and \ref{fig:Fig2}(b), there is a well pronounced difference between the fitted $\alpha$ parameters for three analyzed excitons, beyond experimental error. The difference between the two neutral dark components: $\alpha_{\textrm{X}^\textrm{D}}$=0.23/T$^2$ versus $\alpha_{\textrm{X}^\textrm{G}}$=0.15/T$^2$, can be explained by the thermal occupation of these two states at $T$=10~K corresponding to our experimental conditions. The calculated population ratio is $e^{{-\delta}/{k_B T}}$=0.54, which is in reasonable agreement with the measured ratio $\alpha_{\textrm{X}^\textrm{G}} / \alpha_{\textrm{X}^\textrm{D}}$=0.65. 

The brightening rate of the dark trion with respect to dark excitons is much faster with  $\alpha_{\textrm{T}^\textrm{D}}$$\sim24$$\alpha_{\textrm{X}^\textrm{G}}$. This large difference may be understood considering  the finite free electron concentration of 10$^{11}$~cm$^{-2}$ in the WS$_2$ ML~\cite{kapuscinski2020}. The formation of trions and of dark trions is then favored compared to neutral excitons and this is reflected in the much more pronounced brightening effect for T$^\textrm{D}$ than for X$^\textrm{D/G}$. Note that in our previous studies devoted to the dark excitonic complexes in the WSe$_2$ ML \cite{Molas2019Dark}, we observed the opposite situation, $i.e.$ the brightening of the X$^\textrm{D/G}$ line was significantly larger as compared to the one of the charged complex. This may be understood if the doping level of the WSe$_2$ ML was close to the neutrality point, which favours the creation of the neutral dark excitons.  The obtained results highlight a role of doping in MLs on the brightening of the neutral and charged dark exciton emission in the in-plane magnetic field.

\begin{figure}[!t]
	\centering
	\includegraphics[width=1\linewidth]{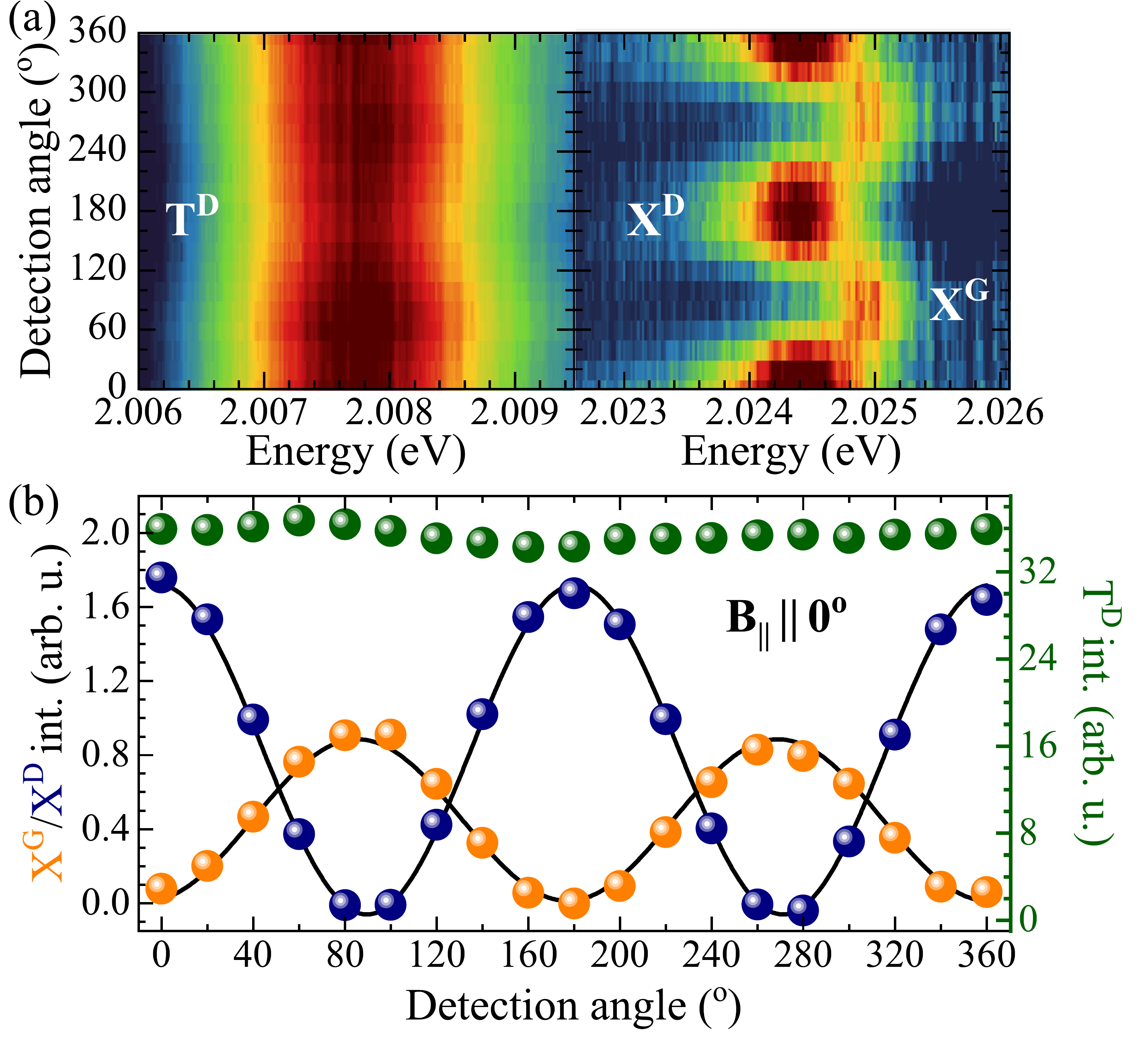}%
	\caption{(a) False-color map of the charged (left panel) and neutral (right panel) dark exciton emissions as a function of the detection linear polarization angle measured at $B_{||}$=10~T. (b) The integrated intensities of the luminescence peaks  measured at $B_{||}$=10 T as a function of the polarization. Solid black curves represent sine fits.}
	\label{fig:aniso}
\end{figure}

The polarization properties of magnetically brightened neutral and charged dark excitons were also analysed (see Fig.~\ref{fig:aniso}(a)). The integrated PL intensity for each peak measured at $B_{||}$=10 T as a function of the linear polarization is shown in Fig.~\ref{fig:aniso}(b). The detection angle is measured between the polarizer axis and the direction of the $B_{||}$ field. The X$^\textrm{G}$ and X$^\textrm{D}$ intensities are expected to follow the $I(\phi_d)\sim I_0 \sin^2(\phi_d + \theta)$ dependence, where $\phi_d$ represents the detection angle, while $I_0$ and $\theta$ are fitting parameters. Our measurements show that the neutral dark (grey) exciton at $B_{||}$=10~T is linearly polarized along (perpendicularly to) the direction of the in-plane magnetic field, in line with our previous study of magnetic brightening in a monolayer of WSe$_2$~\cite{Molas2019Dark}. In contrast to dark excitons, the emission from dark trions do not show sizable linear polarization.

\begin{figure}[t]
	\centering
	\includegraphics[width=1\linewidth]{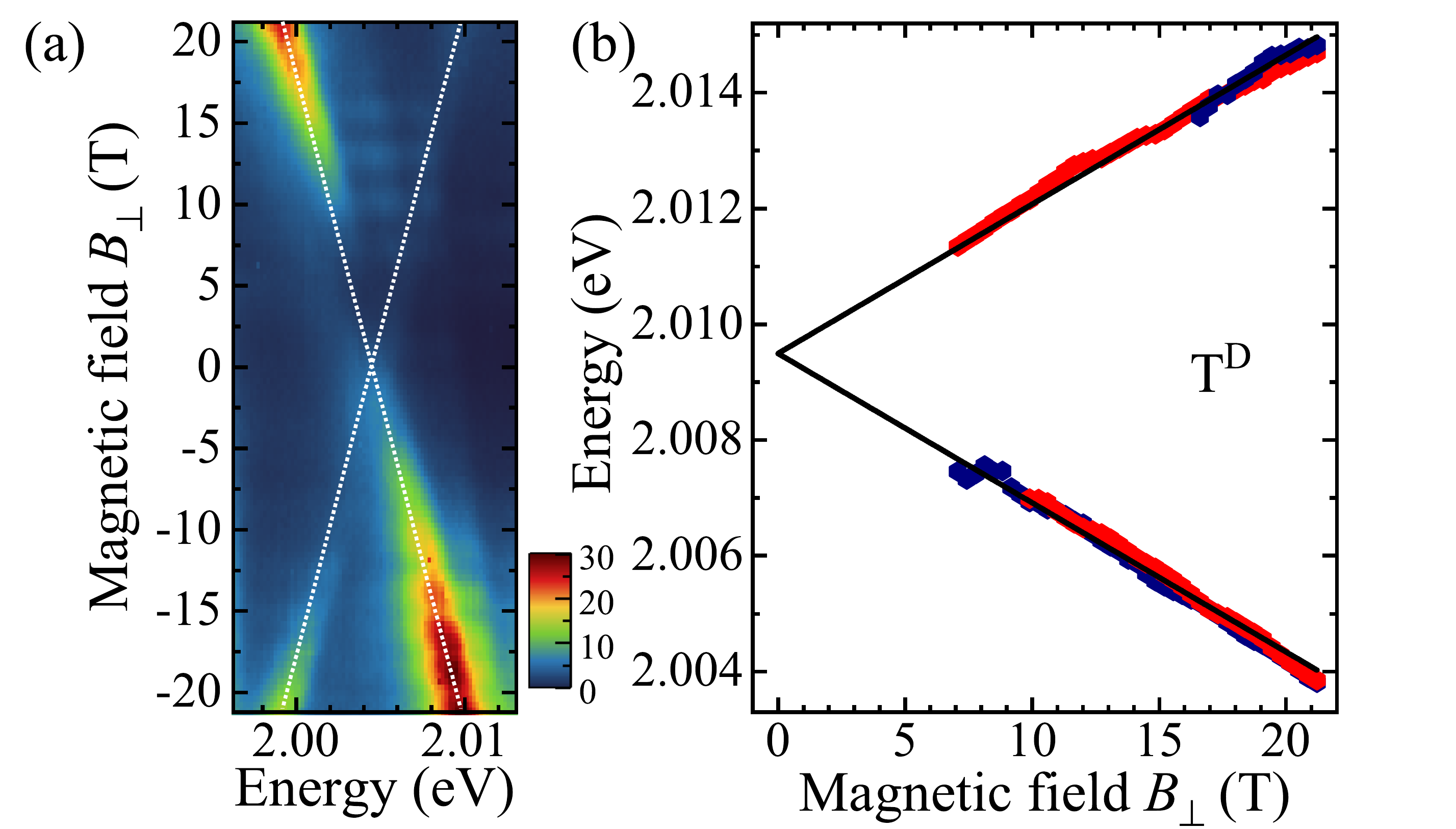}%
	\caption{a) False-color map of the PL response as a function of out-of-plane component ($B_\perp$) of the applied tilted magnetic field. Note that the positive and negative values of magnetic fields correspond to $\sigma^\pm$ polarizations of detection. The intensity scale is normalized to subtract the Faraday effect affecting the experimental data. White dashed superimposed on the investigated transitions are guides to the eyes. (b) Transition energies of the $\sigma^{+/-}$ (red/blue points) components of the T$^\textrm{D}$ line as a function of the out-of-plane magnetic field. The solid lines represent fits according to the equation described in the text.}
	\label{fig:T_dark}
\end{figure}

In order to describe the T$^\textrm{D}$ polarization hallmarks, we analyse the observed sine-square intensity profiles of neutral dark excitons presented in Fig.~\ref{fig:aniso}(b). The evolution results from two effects: i) the coupling between dark and bright excitons of the same valley by an in-plane magnetic field; and ii) the coupling of dark excitons of opposite valleys by short-range exchange interaction. The first effect allows to observe dark excitons by transferring part of the optical activity from bright to dark excitons through the mixing of the the two spin states of the conduction band. Then, dark excitons from K$^\pm$ valleys get the ability to decay radiatively by emitting $\sigma^\pm$ polarized photons in the direction, perpendicular to the monolayer's plane. The exchange interaction mixes the "valley" dark exciton states into new ''grey'' and ''dark'' states \cite{Slobodeniuk2016}. The new states are no more degenerated in energy, are split by the exchange interaction \cite{WangMarie2017}, and are orthogonal to each other. As a result, dark excitons recombine by emitting linearly polarized light with orthogonal polarization. The linear polarization results in the sine-square profile of the PL intensity versus the detection angle, while the orthogonality is responsible for the $\pi/2$ phase shift of corresponding curves. The lowest energy dark exciton state is more populated than the grey exciton state and hence emits more photons~\cite{Molas2019Dark}. The difference between the corresponding intensities becomes significant at low temperatures due to Boltzmann factor $\exp(-\delta/k_BT)$, as it is observed in our study, see Figs~\ref{fig:Fig2} and \ref{fig:aniso}. Note that in the absence of the exchange interaction both sine-square curves would be characterised by the same intensity profile and energy. In such case the valley dark excitonic doublet would remain doubly degenerate and the PL intensity of dark exciton line would become angle-independent, since $\sin^2(\phi_d)+\sin^2(\phi_d+\pi/2)=1$. In fact, such a situation is realized in the case of negative dark trions. They also form the valley doublet, which is brightened by in-plane magnetic field. However, since the dark trions from opposite valleys are not influenced by exchange interaction \cite{Courtade2017}, their emission intensity is not polarized, as it was observed in the experiment (see green points in Fig.~\ref{fig:aniso}(b)). In addition, it is expected that dark trions are characterized by an out-of-plane exciton dipole momentum, which leads to the reported in-plane emission at 55~meV below the X$^\textrm{B}$ in WS$_2$ monolayer \cite{WangMarie2017}.

To determine the dark trion g-factor, it is necessary to apply both an in-plane magnetic field allowing for its direct observation in a photoluminescence experiment and, at the same time, to apply a magnetic field perpendicular to the layer plane to couple to its magnetic moment. This is achieved in our experiment by tilting the layer by 45$^\circ$ with respect to the magnetic field direction. Fig.~\ref{fig:T_dark}(a) shows the evolution of the photoluminescence response as a function of the out-of-plane component of the magnetic field ($B_\perp$), in the form of colour-coded map. The T$^\textrm{D}$ emission grows due to the in-plane field and additionally splits into components due to the valley Zeeman effect. The extracted emission energies of the negative dark trion are presented in the Fig.~\ref{fig:T_dark}(b). The energy evolutions ($E(B)$) in external out-of-plane magnetic fields ($B_\perp$) can be described as $E(B)=E_0\pm\frac{1}{2} g \mu_\mathrm{B} B_\perp$, where $E_0$ is the energy of the transition at zero field, $g$ denotes the g-factor of the considered excitonic complex and $\mu_{B}$ is the Bohr magneton. The black solid lines in Fig.~\ref{fig:T_dark}(b) are fits to our experimental data using this equation. We find that the T$^\textrm{D}$ g-factor in the studied WS$_2$ monolayer is on the order of -8.9, which is very similar to the reported g-factors for dark trions in WSe$_2$ MLs~\cite{Liu2019, Li2019Trion, Liu2019Replica, he2020valley}, about two times bigger than the g-factors of the X$^\textrm{B}$, T$^\textrm{S}$ and T$^\textrm{T}$ lines (for details see Supplementary Material (SM)).

\begin{figure}[t]
	\centering
	\includegraphics[width=1\linewidth]{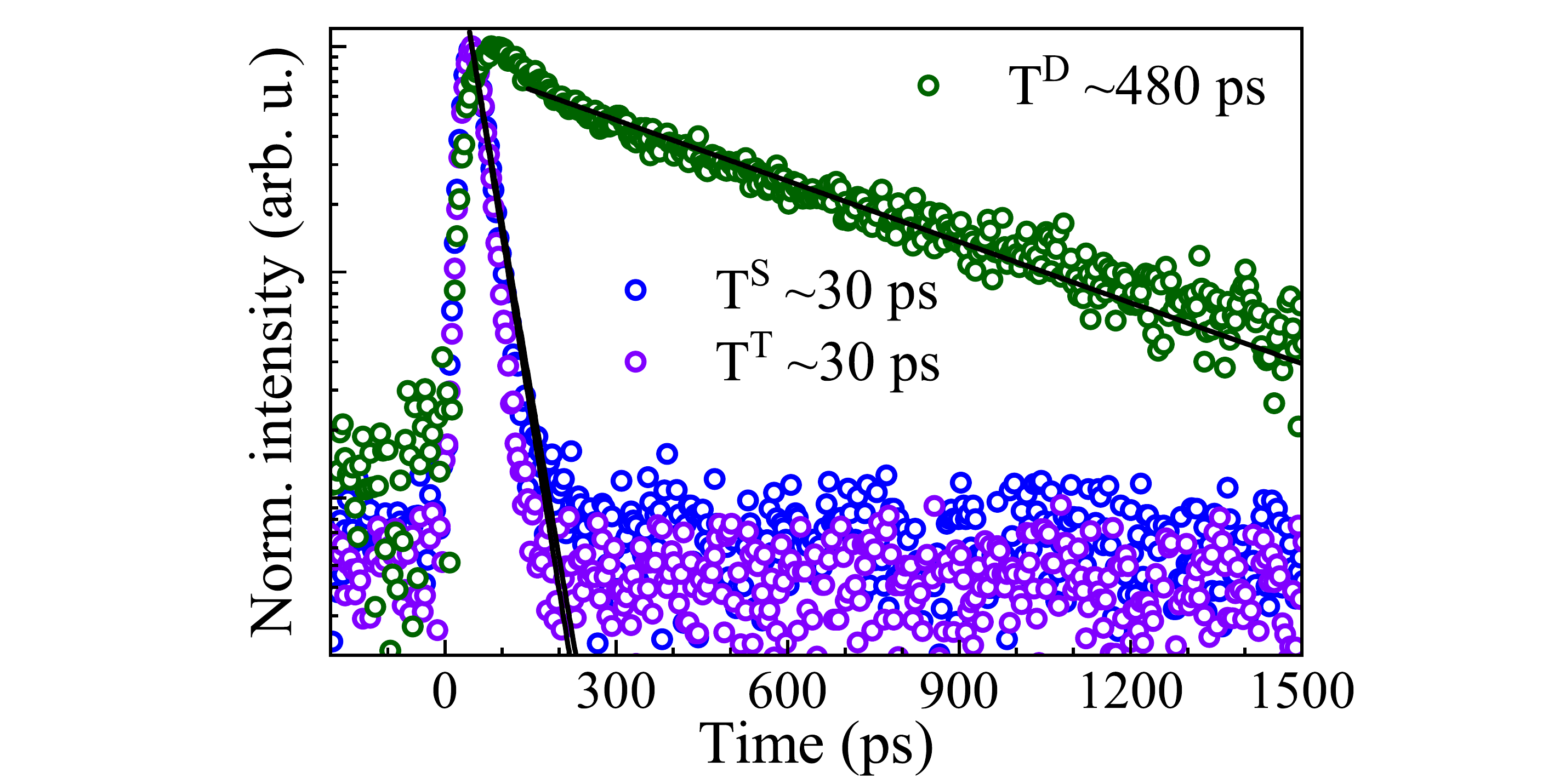}%
		\caption{Low-temperature normalized time-resolved PL of the T$^\textrm{T}$, T$^\textrm{S}$ and T$^\textrm{D}$ lines measured on a WS$_2$ ML encapsulated in hBN flakes at in-plane magnetic field $B_{||}$=10 T.}
	\label{fig:lifetime}
\end{figure}

To get more information on the dark trion, we have performed time-resolved measurements at in-plane magnetic field of $B_{||}$=10 T. Fig.~\ref{fig:lifetime} displays the normalized time-resolved traces of the bright singlet (T$^\textrm{S}$) and triplet (T$^\textrm{T}$), together with the dark (T$^\textrm{D}$) charged excitons. Monoexponential fits of these traces show that the decay time of the two bright trions is of 30 ps, in line with previous studies for monolayers of other S-TMD\cite{Li2019Trion,Liu2019}.  Dark trions are characterized by a much slower decay time, reaching values close to 0.5 ns. The obtained decay times of the bright and dark trions are in good agreement with previously reported values for other S-TMD monolayer~\cite{Liu2019,Li2019Trion}.

\section{Methods \label{methods}}

The studied sample is composed of WS$_2$ ML encapsulated in hBN flakes and supported by a bare Si substrate. The structure was obtained by two-stage polydimethylsiloxane (PDMS)-based \cite{Gomez2014} mechanical exfoliation of WS$_2$ and hBN bulk crystals. A bottom layer of hBN in the hBN/WS$_2$/hBN heterostructure was created in the course of a non-deterministic exfoliation. The assembly of the hBN/WS$_2$/hBN heterostructure was realized via succesive dry transfers of WS$_2$ ML and capping hBN flake from PDMS stamps onto the bottom hBN layer.

Low-temperature micro-magneto-PL experiments are performed in the Voigt, Faraday and tilted geometries, $i.e.$ magnetic field oriented parallel, perpendicular, and 45$^\circ$ with respect to ML's plane. Measurements (spatial resolution $\sim$2~$\mu$m) were carried out with the aid of two systems: a split-coil superconducting magnet and a resistive solenoid producing fields up to 10~T and 30~T using a free-beam-optics arrangement and an optical-fiber-based insert, respectively. The sample was placed on top of a $x$-$y$-$z$ piezo-stage kept at $T$=10~K or $T$=4.2~K and was excited using a laser diode with 532~nm or 515~nm wavelength (2.33 eV or 2.41 eV photon energy). The emitted light was dispersed with a 0.5~m long monochromator and detected with a charge coupled device (CCD) camera. The linear polarizations of the emissions in the Voigt geometry were analyzed using a set of polarizers and a half wave plate placed directly in front of the spectrometer. In the case of the Faraday and tilted-field configurations, the combination of a quarter wave plate and a linear polarizer placed in the insert were used to analyse the circular polarization of signals (the measurements were performed with a fixed circular polarization, whereas reversing the direction of magnetic field yields the information corresponding to the other polarization component due to time-reversal symmetry). For time-resolved measurements, a femtosecond pulsed laser with excitation at 570 nm (2.18 eV photon energy) from frequency-doubled Ti:Sapphire Coherent Mira-OPO laser system operated at a repetition rate of 76 MHz and a synchroscan Hamamatsu streak camera were used correspondingly for excitation and detection. Note that the excitation power for experiments performed in magnetic fields up to 30~T and with time resolution was adjusted based on the comparison of the measured PL spectrum and the one obtained under excitation of laser with 532~nm in measurement in fields up to 10~T.

\section{Summary}
We have presented a photoluminescence investigation of dark exciton complexes in a monolayer of WS$_2$ encapsulated in hBN. Based on polarization resolved measurements, we have identified both the dark and the grey excitons which are brightened by the in-plane component of the magnetic field. The dark trion is also observed in this magneto-brightening experiment. In contrast to dark excitons, dark trions do not show any fine structure nor linear polarization due to the absence of coupling between dark trions in the two valleys. Time resolved measurements indicate that the decay time of dark trions is close to 0.5 ns, which is more than two orders of magnitude larger than that of bright trions in the same monolayer. This study also indicates that monolayer of WS$_2$ encapsulated in hBN is naturally $n$-doped as revealed by the much higher brightening rate of the dark trion with respect to dark excitons, in contrast to previous results obtained in WSe$_2$.

\section*{Acknowledgements}
The work has been supported by the the National Science Centre, Poland (grants no. 2017/27/B/ST3/00205, 2017/27/N/ST3/01612 and 2018/31/B/ST3/02111), EU Graphene Flagship project (no. 785219), the ATOMOPTO project (TEAM programme of the Foundation for Polish Science, co-financed by the EU within the ERD-Fund), the Nano fab facility of the Institut N\'eel, CNRS UGA, and the LNCMI-CNRS, a member of the European Magnetic Field Laboratory (EMFL). P. K.  kindly acknowledges the National Science Centre, Poland (grant no. 2016/23/G/ST3/04114) for financial support for his PhD. M. B. acknowledges the financial support of the Ministry of Education, Youth and Sports of the Czech Republic under the project CEITEC 2020 (Grant No. LQ1601). K.W. and T.T. acknowledge support from the Elemental Strategy Initiative conducted by the MEXT, Japan, (grant no. JPMXP0112101001), JSPS KAKENHI (grant no. JP20H00354), and the CREST (JPMJCR15F3), JST.

\bibliographystyle{apsrev4-1}
\bibliography{Dark_WS2}

\newpage
\onecolumngrid
\setcounter{figure}{0}
\setcounter{section}{0}
\renewcommand{\thefigure}{S\arabic{figure}}
\renewcommand{\thesection}{S\arabic{section}}

\begin{center}
	{\large{ {\bf Supplemental Material: \\ Neutral and charged dark excitons in monolayer WS$_2$}}}
	\vskip0.5\baselineskip{M. Zinkiewicz,{$^{1}$} A. O. Slobodeniuk,{$^{2}$} T. Kazimierczuk,{$^{1}$} P. Kapu\'sci\'nski,{$^{3,4}$} K. Oreszczuk,{$^{1}$} M. Grzeszczyk,{$^{1}$} M.~Bartos,{$^{3,5}$} K. Nogajewski,{$^{1}$} K. Watanabe,{$^{6}$} T. Taniguchi,{$^{7}$} C. Faugeras,{$^{3}$} P. Kossacki,{$^{1}$} M. Potemski,{$^{1,3}$} A.~Babi\'nski,{$^{1}$} and M. R. Molas {$^{1}$}}
	\vskip0.5\baselineskip{\em$^{1}$ Institute of Experimental Physics, Faculty of Physics, University of Warsaw, ul. Pasteura 5, 02-093 Warsaw, Poland \\$^{2}$ Department of Condensed Matter Physics, Faculty of Mathematics and Physics, Charles University in Prague, Ke Karlovu 5, Praha 2 CZ-121 16, Czech Republic \\$^{3}$ Laboratoire National des Champs Magn\'etiques Intenses, CNRS-UGA-UPS-INSA-EMFL, 25, avenue des Martyrs, 38042 Grenoble, France \\$^{4}$ Department of Experimental Physics, Faculty of Fundamental Problems of Technology, Wroc\l{}aw University of Science and Technology, ul. Wybrze\.ze Wyspia\'nskiego 27, 50-370 Wroc\l{}aw, Poland \\$^{5}$ Central European Institute of Technology, Brno University of Technology,  Purky\v{n}ova 656/123, 612 00 Brno, Czech Republic National Institute for Materials Science, 1-1 Namiki, Tsukuba 305-0044, Japan \\$^{6}$ Research Center for Functional Materials, National Institute for Materials Science, 1-1 Namiki, Tsukuba 305-0044, Japan \\$^{7}$ International Center for Materials Nanoarchitectonics, National Institute for Materials Science, 1-1 Namiki, Tsukuba 305-0044, Japan}
\end{center}

\begin{figure}[b]
	\centering
	\includegraphics[width=0.5\linewidth]{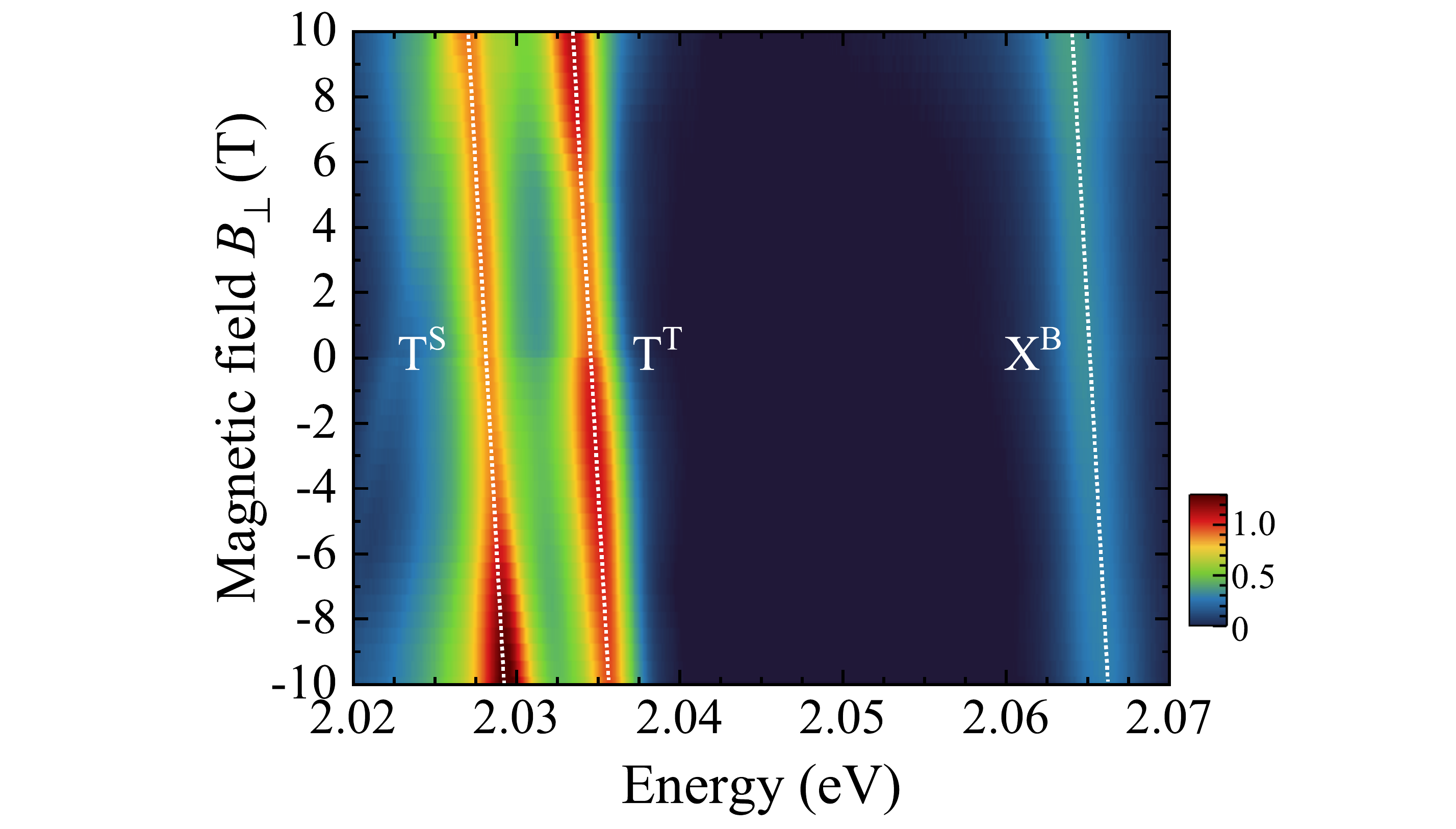}%
	\caption{False-color map of the PL response as a function of $B_\perp$. Note that the positive and negative values of magnetic fields correspond to $\sigma^\pm$ polarizations of detection. The intensity scale is logarithmic. White dashed lines superimposed on the investigated transitions are guides to the eyes.}
	\label{fig:mapa1}
\end{figure}

\section{g-factors of bright excitonic complexes}
\begin{figure}[t]
	\centering
	\includegraphics[width=0.5\linewidth]{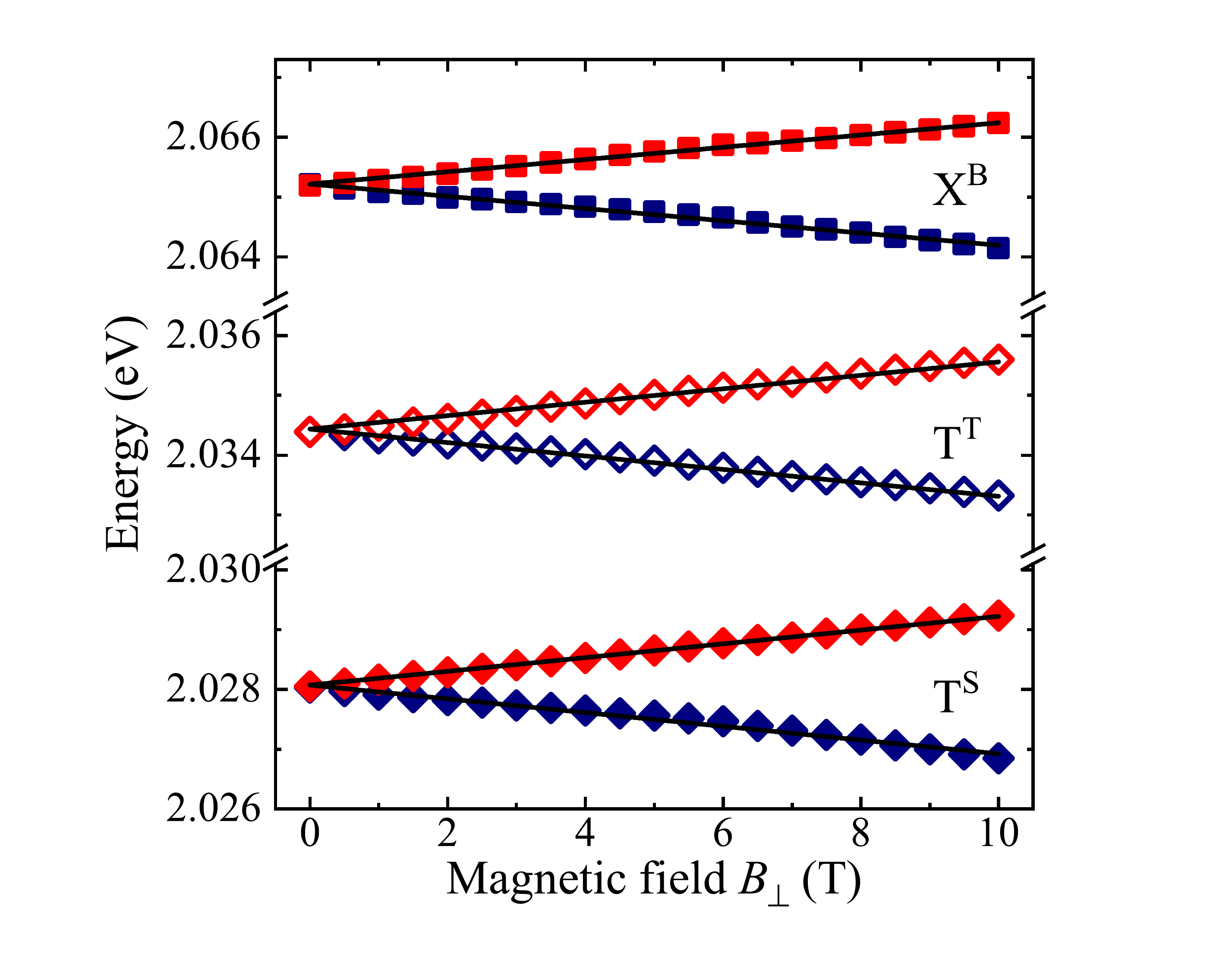}%
	\caption{Transition energies of the $\sigma^{+/-}$ (blue/red points) components of the~X$^\textrm{B}$, T$^\textrm{S}$ and T$^\textrm{T}$ lines as a function of the~out-of-plane magnetic field. The solid lines represent fits according to Eq.~\ref{eq:zeeman}.}
	\label{fig:Zeeman}
\end{figure}

To get more information on the properties of the bright excitons emissions in the studied WS$_2$ monolayer, we performed the magneto-photoluminescence experiment in magnetic fields up to 10~T oriented perpendicular to ML's plane. Fig.~\ref{fig:mapa1} illustrates the measured PL spectra as a function of magnetic fields in the form of colour-coded map. Upon application of an out-of-plane magnetic field, the excitonic emissions split into two circularly polarized components due to the excitonic Zeeman effect~\cite{koperski2019SM}. Their energies evolutions ($E(B)$) in external out-of-plane magnetic fields ($B_\perp$) can be described as:
\begin{equation}
E(B)=E_0\pm\frac{1}{2} g \mu_\mathrm{B} B_\perp,
\label{eq:zeeman}
\end{equation}
where $E_0$ is the energy of the transition at zero field, $g$ denotes the g-factor of the considered excitonic complex and $\mu_{B}$ is the Bohr magneton. The results of fitting to the experimental results denoted by red and blue points, are presented in Fig.~\ref{fig:Zeeman} as solid black curves. We found that the g-factors for the bright exciton (X$^\textrm{B}$) and singlet (T$^\textrm{S}$) and triplet (T$^\textrm{T}$) states of negative trions are of about -3.5, -4.0 and -3.9, respectively. The obtained values are in in very close agreement to previous measurements on WS$_2$ monolayer~\cite{Stier2016SM, Plechinger2016SM, Arora2016SM, koperski2019SM, kapuscinski2020SM}. Summarizing, the g-factors of the bright excitons, $i.e.$ complexes for which recombining an electron and a hole posses the same sign of the spin, are very close to the theoretically predicted value of 4~\cite{koperski2019SM}.

\end{document}